\documentclass[aps,prc,twocolumn,superscriptaddress]{revtex4-1}

\usepackage{xtab,booktabs}
\usepackage{color}
\usepackage{amsmath}
\usepackage{amssymb}
\usepackage{graphicx}
\usepackage{lineno}

\begin{document}



\title{Systematic trends of neutron skin thickness versus relative neutron excess}



\author{J. T. Zhang}
\affiliation{Joint Department for Nuclear Physics, Lanzhou University and Institute of Modern Physics, Chinese Academy of Sciences, Lanzhou 730000, China}
\affiliation{School of Nuclear Science and Technology, Lanzhou University, Lanzhou 730000, China}
\affiliation{Institute of Modern Physics, Chinese Academy of Science, Lanzhou 730000, China}
\author{X. L. Tu}
\email{tuxiaolin@impcas.ac.cn}
\affiliation{Institute of Modern Physics, Chinese Academy of Science, Lanzhou 730000, China}
\affiliation{School of Nuclear Science and Technology, University of Chinese Academy of Sciences, Beijing 100049, China}
\affiliation{Max-Planck-Institut  f\"ur Kernphysik, Saupfercheckweg 1, 69117 Heidelberg, Germany}
\author{P. Sarriguren}
\affiliation{Instituto de Estructura de la Materia, CSIC, Serrano 123, E-28006 Madrid, Spain}

\author{K. Yue}
\email{yueke@impcas.ac.cn}
\affiliation{Institute of Modern Physics, Chinese Academy of Science, Lanzhou 730000, China}

\author{Q. Zeng}
\affiliation{Institute of Modern Physics, Chinese Academy of Science, Lanzhou 730000, China}
\affiliation{Engineering Research Center of Nuclear Technology Application, East China University of Technology, Nanchang 330013,
China}
\author{Z. Y. Sun}
\author{M. Wang}
\author{Y. H. Zhang}
\author{X. H. Zhou}
\affiliation{Institute of Modern Physics, Chinese Academy of Science, Lanzhou 730000, China}
\author{Yu. A. Litvinov}
\affiliation{Institute of Modern Physics, Chinese Academy of Science, Lanzhou 730000, China}
\affiliation{GSI Helmholtzzentrum f\"ur Schwerionenforschung, D-64291 Darmstadt, Germany}


\date{\today}

\begin{abstract}
Available experimental neutron skin thicknesses of even-even stable Ca, Ni, Sn, Pb, and Cd isotopes are evaluated, and
separate trends of neutron skin thickness versus relative neutron excess $\delta=(N-Z)/A$ are firstly observed for different isotopic chains.
This phenomenon is quantitatively reproduced by the deformed Skyrme Hartree-Fock $+$ BCS model with SLy4 force.
\end{abstract}

\pacs{}

\maketitle
\section{Introduction}
Nucleus is a quantum many-body system consisting of neutrons and protons.
The root-mean-square (rms) radii of neutron and proton, which characterize the spatial matter density distributions of neutrons and protons in a nucleus, are fundamental properties of the nucleus~\cite{Tanihata13,Egelhof01}.
Nuclear charge rms radii determined by different experimental methods were evaluated in~\cite{Angeli13}, and the rms values with precisions better than 0.01~fm were reported.
These valuable quantities are usually used to test and constrain microscopic theories, for instance, by which odd-even staggering of charge radii of exotic copper isotopes~\cite{Groote20} and shape-staggering effects in mercury isotopes~\cite{Marsh18} have been well explained by dedicated theoretical calculations.
It is noted that the proton distribution rms radius can be deduced from nuclear charge rms radius~\cite{Terashima08}, and therefore, the precision of proton rms radius is also high.
Different from the proton rms radii which are related to the well-known electromagnetic interaction, the determinations of neutron distribution rms radii are model-dependent and are much more complicated~\cite{Krasznahorkay04}.
Consequently, uncertainties of extracted neutron rms radii are relatively large and depend on the model uncertainties.
Nonetheless, these quantities are still sensitive for probing nuclear structure~\cite{Tanihata85,Bagchi19}.

Neutron skin thicknesses, $\Delta r_{np}$, defined as the difference of neutron and proton rms radii of a nucleus, are indispensable in nuclear reaction and nuclear astrophysics researches~\cite{Li08,Fattoyev18}.  
A variety of microscopic~\cite{Sarriguren07,Warda98-2,Seif15,Furnstahl02} and macroscopic~\cite{Myers80,Pethick96,Warda09,Iida04} models were developed to describe $\Delta r_{np}$.  
We emphasize that theories play a decisive role in constraining the parameters of the equation of state (EOS) of isospin asymmetric nuclear matter~\cite{Li08}.
For example, a linear correlation between the $\Delta r_{np}$ and the slope of symmetry energy at the saturation density was deduced through microscopic mean-field calculations~\cite{Li08,Brown00}.

Experimentally, by using the $\Delta r_{np}$ data with large uncertainties, a linear dependence of $\Delta r_{np}$ on the relative neutron excess, $\delta=(N-Z)/A$, was reported with a fitting goodness ($\chi^2$) of 0.6, see Fig.~4 in~\cite{Trzcinska01}.
This result has extensively been used to constrain theories~\cite{Warda09,Centelles09,Warda10,Bertulani12,Kumar18,Thiel19} and to predict the nuclear $\Delta r_{np}$ values as well.
For instance, the predicted $\Delta r_{np}$ of  $^{133}$Cs is employed in the study of atomic parity violation for testing the standard model of elementary particle physics at low energies~\cite{Derevianko01}.

We know that the nuclear structure is reflected in the nucleon distribution radii~\cite{Bagchi19,Suzuki98,Tanaka20}. 
For example, larger neutron and proton radii were observed in deformed nuclei~\cite{Tanihata85,Rodriguez10}. 
Such deformation-related effects may alter the assumed linear behavior of nucleon radii for the isotopic chains~\cite{Trzcinska01}.
If the uncertainties of neutron skin thicknesses are improved, what can be observed on the linear trend reported in~\cite{Trzcinska01}?
In this work, available $\Delta r_{np}$ data of even-even stable Ca, Ni, Sn, Pb, and Cd isotopes are evaluated in order to study the systematic behavior of $\Delta r_{np}$ versus $\delta$.  

\section{Experimental data evaluation}
In order to study the systematic behavior of $\Delta r_{np}$ along with $\delta$ for different isotopic chains, $\Delta r_{np}$ of even-even stable Ca, Ni, Sn, Pb, and Cd isotopes are evaluated.
Taking the rms radius, $\left \langle r_{m}^2 \right \rangle^{1/2}$, of point-matter distribution, the $\Delta r_{np}$ is deduced via~\cite{Chaumeaux78}
\begin{equation}
\begin{aligned}
\Delta r_{np}=\left \langle r_{n}^2 \right \rangle ^{1/2}-\left \langle r_{p}^2 \right \rangle ^{1/2}\quad,\\
 \left \langle r_{n}^2 \right \rangle^{1/2}=\sqrt{\frac{A}{N}\left \langle r_{m}^2 \right \rangle-\frac{Z}{N}\left \langle r_{p}^2 \right \rangle}\quad,
 \end{aligned}
  \end{equation}
where $\left \langle r_{n}^2 \right \rangle^{1/2}$ and $\left \langle r_{p}^2 \right \rangle^{1/2}$ are point-neutron and point-proton distribution rms radii, respectively.
The point-proton rms radius is related to the nuclear charge rms radius as $\left \langle r_{p}^2 \right \rangle=\left \langle r_{ch}^2 \right \rangle-0.64$ in our analysis.
Small corrections on $\left \langle r_{p}^2 \right \rangle^{1/2}$ resulting from spin-orbit term etc. were taken into account in some adopted data~\cite{Brown07}.
Compared to the uncertainties of neutron distribution rms radii, the difference are very small and thus the corrections are neglected in the present analysis.
Since the folded matter distribution rms radii, $\left \langle  \widetilde r_{m}^2 \right \rangle^{1/2}$, such as for the Cd isotopes~\cite{Miller81}, contain the finite size of the nucleon, the corresponding point-matter rms radii are deduced via $\left \langle r_{m}^2 \right \rangle ^{1/2}=\sqrt{\left \langle \widetilde r_{m}^2 \right \rangle-0.64}$~\cite{Chaumeaux78}.
The evaluated neutron skin thicknesses are weighted averages.
Table~\ref{tab1} in Appendix lists the $\Delta r_{np}$ values evaluated in this work. The precisions of $\Delta r_{np}$ have been improved.

\section{Systematic trends and discussions}
A linear relationship between $\Delta r_{np}$ and $\delta$ was reported in~\cite{Trzcinska01}, but the used data have large statistical errors.
The $\Delta r_{np}$ of even-even stable Ca, Ni, Sn, Pb isotopes were determined by many experiments, and consequently, the evaluated $\Delta r_{np}$ values have less uncertainties as shown in Table~\ref{tab1} of Appendix.
Moreover, ground states of even-even nuclei with magic proton numbers have spherical shapes, and hence the influence of deformation on the neutron skin thickness is minimal. 
The evaluated neutron skin thicknesses for even-even Ca, Ni, Sn, Pb isotopes are thought to be reliable to study the systematic correlations between $\Delta r_{np}$ and $\delta$. 

\begin{figure}[h!]
\begin{center}
\includegraphics*[width=7.cm]{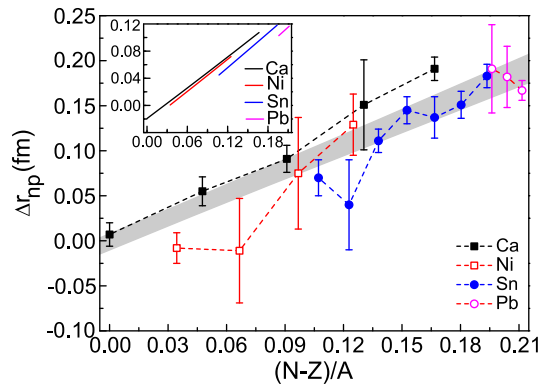}
\caption{(Color online). Neutron skin thicknesses, $\Delta r_{np}$, of even-even stable Ca, Ni, Sn, and Pb isotopes as a function of relative neutron excess $\delta=(N-Z)/A$. 
The gray area shows the global linear fit with the corresponding uncertainties.
The solid lines in the inset represent the macroscopic model calculations~\cite{Iida04}. }
\label{fig1}
\end{center}
\end{figure}

The correlation between $\Delta r_{np}$ and $\delta$ is shown in Fig.~\ref{fig1}. Although an overall linear relationship of $\Delta r_{np}$ versus $\delta$ is observed, 
the normalized {\it chi} value from a linear fit to all the data in Fig.~\ref{fig1} is $\chi_n=$1.32. This value is apparently outside of the expected 1$\sigma$ range of $\chi_n=1\pm0.17$. 
Thanks to the improved precisions of $\Delta r_{np}$, the curves of $\Delta r_{np}$ versus $\delta$ for Ca, Ni, Sn, and Pb isotopic chains are separated from each other, as demonstrated in Fig.~\ref{fig1}. We note that the overall linear relationship reported in~\cite{Trzcinska01} is in fact composed of several individual curves for different isotopic chains.

\begin{figure}[h!]
\begin{center}
\includegraphics*[width=7.cm]{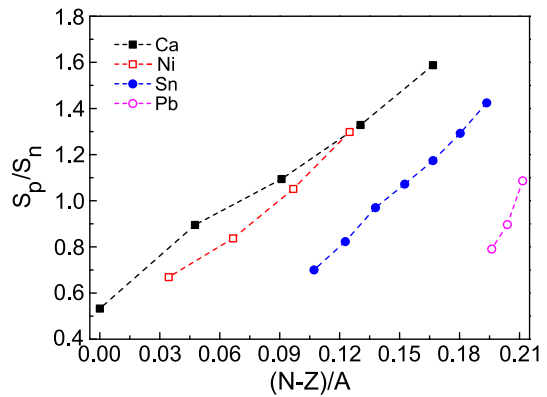}
\caption{(Color online). Ratios of proton to neutron separation energies, $S_p/S_n$, as a function of relative neutron excess $\delta=(N-Z)/A$ for the Ca, Ni, Sn and Pb isotopic chains.}
\label{fig2}
\end{center}
\end{figure}

It is known that there is a strong correlation between $\Delta r_{np}$ and nucleon separation energy~\cite{Ozawa01}. 
In general, a larger proton separation energy, $S_p$, results in a larger neutron skin thickness, and a larger neutron separation energy, $S_n$, leads to a smaller neutron skin thickness~\cite{Ozawa01}. 
Due to different separation energies, $\Delta r_{np}$ would be distinguishable for nuclides with the same relative neutron excess. 
Figure~\ref{fig2} shows the $S_p/S_n$ ratios as a function of $\delta$. One see that the $S_p/S_n\sim \delta$ plot has a similar pattern as that of $\Delta r_{np}\sim \delta$ in Fig.~\ref{fig1}.

Neutron skin thickness can be calculated by both microscopic\cite{Sarriguren07,Warda98-2,Seif15,Furnstahl02} and macroscopic~\cite{Myers80,Pethick96,Warda09,Iida04} models. In the present work, the deformed Hartree-Fock (HF) plus BCS method based on the SLy4 Skyrme force are used to calculate the neutron skin thickness. Details on the deformed Skyrme HF+BCS model are referred to~\cite{Sarriguren07,Sarriguren19}.

\begin{figure}[h!]
\begin{center}
\includegraphics*[width=7.cm]{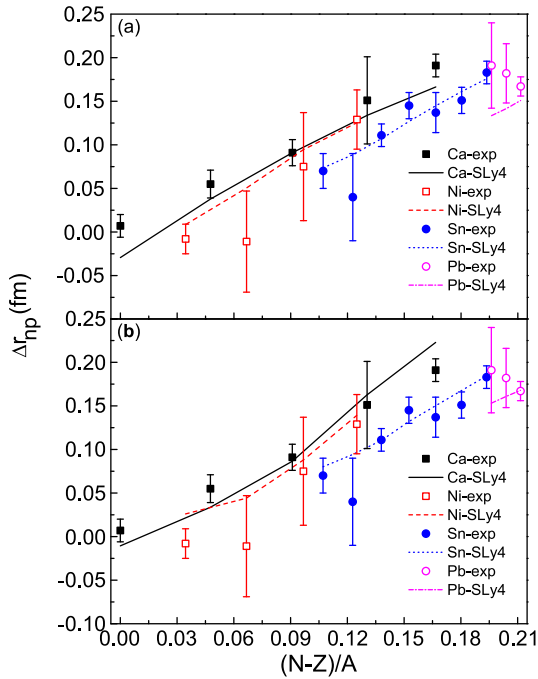}
\caption{(Color online). Comparison of evaluated $\Delta r_{np}$ values and the those calculated by the HF+BCS method.
(a) Both neutron and charge rms radii are taken from the HF+BCS calculations with the SLy4 Skyrme force~\cite{Sarriguren07}.
(b) Same as (a) but using the experimental charge rms radii from~\cite{Angeli13}.}
\label{fig3}
\end{center}
\end{figure}

Theoretical neutron skin thickness is extracted via $\Delta r_{np}=\left \langle r_{n}^2 \right \rangle^{1/2}-\sqrt{\left \langle r_{ch}^2 \right \rangle-0.64}$,
where neutron and charge rms radii are calculated by the HF+BCS~\cite{Sarriguren07}.
Figure~\ref{fig3}(a) shows the comparison of evaluated and calculated $\Delta r_{np}$ values.
One see that the evaluated $\Delta r_{np}$ data are practically reproduced by the theoretical calculations, and the theory yields separate $\Delta r_{np}\sim \delta$ curves for different isotopic chains. 
On the other hand, we re-calculated the $\Delta r_{np}$ values using the experimental charge rms radii and the theoretical neutron radii. The results are given in Fig.~\ref{fig3}(b). It is worth noting that both calculations yield consistent results, and noticeably a better agreement is achieved for Pb isotopic chain by using the experimental nuclear charge radii.

We would like to point out that the unevaluated experimental $\Delta r_{np}$ values for Sn isotopes locate in-between the SLy4 and RMF predictions (see Fig.~4 in~\cite{Sarriguren07}). This indicates that the theoretical models can not be effectively constrained by the unevaluated data. However, as shown in Fig.~\ref{fig3}, the high precision of our evaluated $\Delta r_{np}$ values makes it possible to constrain the theoretical models.

\begin{figure}[h!]
\begin{center}
\includegraphics*[width=7.cm]{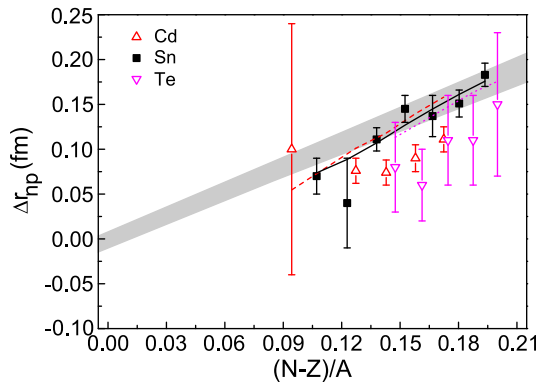}
\caption{(Color online). The neutron skin thicknesses of Cd, Sn, and Te isotopes as a function of the relative neutron excess $(N-Z)/A$. The gray area is same as that in Fig.~\ref{fig1}.
The dashed, solid, and dotted lines represent the theoretical $\Delta r_{np}$ for Cd, Sn, and Te isotopic chains from the HF+BCS calculations with the SLy4 Skyrme force~\cite{Sarriguren07}, respectively.}
\label{fig4}
\end{center}
\end{figure}

The macroscopic compressible liquid-drop model gives a formula of neutron skin thickness expressed as~\cite{Iida04} 
\begin{equation}
\begin{aligned}
&\Delta r_{np}^{b}\simeq\sqrt{\frac{3}{5}}\left[C\left(\delta-\frac{Ze^2}{20R_pE_{s}}\right)\left(1+\frac{3C}{2R_p} \right)^{-1}-\frac{Ze^2}{70E_{s}} \right]\quad,\\
&C=\frac{2\sigma_0}{E_{s}\rho_0}\left(C_{s}+\frac{3L\chi}{K_0} \right)\quad,
\end{aligned}
\end{equation}
where $E_{s}$ denotes the symmetry energy, $L$ the slope of the symmetry energy at saturation density $\rho_0$, and $K_0$ the incompressibility of symmetric nuclear matter. 
$\sigma_0$ and $C_{s}$ represent the coefficients of symmetric matter surface tension and surface-asymmetry, respectively. 
More details are given in~\cite{Iida04}. The separated curves for different isotopic chains can be also obtained by the formula of macroscopic model, see the inset in Fig.~\ref{fig1}.
Let us now discuss the deformed nuclei Cd and Te.
Figure~\ref{fig4} shows $\Delta r_{np}$ of the Cd, Sn, and Te isotopes as a function of $\delta$.
The $\Delta r_{np}$ data for Te isotopes were taken from~\cite{Jastrzebski04}. 
The fitted curve for the spherical nuclei in Fig.~\ref{fig1} and the HF+BCS theoretical calculations~\cite{Sarriguren07} are also shown for comparison.
The theoretical $\Delta r_{np}$ values of Cd and Te are located above and below the calculated curve of the Sn isotopic chain, respectively, and their difference are very small for the Cd, Sn, and Te isotopic chains, see Fig.~\ref{fig4}.
However, the experimental $\Delta r_{np}$ values for the Cd and Te isotopes are generally smaller than the experimental ones of the Sn isotopes. Compared to the global linear fit, systematic lower $\Delta r_{np}$ values for the Te isotopes were also reported in~\cite{klos04}.

If only the contribution due to the quadrupole deformation is considered, the  $\Delta r_{np}$ for the deformed nucleus 
can be related to the proton and neutron radii of an assumed spherical shape via~\cite{Seif15,Suzuki98}
\begin{equation}
\begin{aligned}
&\Delta r_{np}=\left \langle r_n^2 \right \rangle_{def} ^{1/2}-\left \langle r_p^2 \right \rangle_{def} ^{1/2}\\
                       &=\left \langle r_n^2 \right \rangle_{sph} ^{1/2}(1+ \frac{5}{4\pi}\beta_{2,n}^2)^{1/2}-\left \langle r_p^2 \right \rangle_{sph} ^{1/2}(1+ \frac{5}{4\pi}\beta_{2,p}^2)^{1/2}\quad,
\label{e1}
\end{aligned}
\end{equation}
where $\left \langle r_{n(p)}^2 \right \rangle^{1/2}_{def}$ and $\left \langle r_{n(p)}^2 \right \rangle^{1/2}_{sph}$ are the rms radii of the neutron (proton) distribution of the deformed and spherical nuclei, and $\beta_{2,n}$ and $\beta_{2,p}$ are the quadrupole deformation parameters for neutron and proton distributions, respectively.

According to Eq.~(\ref{e1}), a larger neutron skin thickness is expected for a deformed nucleus, assuming nucleus has the same deformation parameters for the neutron and proton distributions.
However, owing to the strong Coulomb repulsion of protons, the calculations in~\cite{Warda98,Barant95} showed that in general the neutron matter distribution is more spherical than the proton matter distribution. 
Hence, the quadrupole deformation for the proton distribution is larger than that for the neutron distribution~\cite{Warda98,Pomorski97,Clement82}. This difference has been observed for Cd and Te isotopes~\cite{Madsen84}. 
As a result, smaller $\Delta r_{np}$ is expected comparing to the corresponding spherical nuclei.
Thus, different deformations for neutron and proton distributions may be a reason for the deviations in Fig.~\ref{fig4}. 

Including the deformed nuclei in the global linear fit, the goodness of the fit becomes evidently worse.
The normalized $\chi_n$ is obtained to be 1.55 from the global fit to the data of the Ca, Ni, Sn, Pb, Cd, and Te isotopes. 
This $\chi_n$ value is significantly outside the expected $1\sigma$ range of $\chi_n=1\pm0.14$. 
Due to large uncertainties, improved experimental data for Cd and Te isotopes are needed to confirm or disprove the present results.

Neutron skin thicknesses play an important role in constraining the EOS parameters. The SLy4 Skyrme force reproduces the separate trends observed in this work. With the SLy4 Skyrme force, the deduced EOS parameters $E_{s}$, $L$ and $K_{0}$ at saturation density are 32.00 MeV, 45.94 MeV, and 229.91 MeV~\cite{Dutra12}, respectively. These parameters are consistent with recent Bayesian analysis by using $\Delta r_{np}$ of Sn isotopes~\cite{Xu20}, also in agreement with various new analyses based on neutron star data since GW170817~\cite{Li21}. However, a value of 106(37) MeV for $L$ was determined recently~\cite{Reed21} by the $\Delta r_{np}$ of 0.283(71) fm for $^{208}$Pb~\cite{Adhikari21}, which was deduced by model-independent parity violation electron scattering. Compared to our evaluated value for  $^{208}$Pb, the deviation is 0.116(72) fm. The $\Delta r_{np}$ of $^{48}$Ca from the CREX experiment would help to clarify the difference~\cite{Horowitz14}. As mentioned in~\cite{Li21}, these interesting tensions inspirit the community to make further researches.

\section{Summary}
Experimental neutron skin thicknesses, $\Delta r_{np}$, for the even-even stable Ca, Ni, Sn, Pb, and Cd isotopes have been evaluated.
Systematic trends of the evaluated $\Delta r_{np}$ as a function of relative neutron excess $\delta$ are investigated.
Separate curves of $\Delta r_{np}$ versus $\delta$ for different isotopic chains are observed from analysis of the evaluated data. This behavior has been practically reproduced by the microscopic and macroscopic models.
Comparing to the experimental data of Sn isotopes, the $\Delta r_{np}$ values of Cd and Te isotopes are systematically smaller. 
This might be understood by taking into account the different deformations of proton and neutron distributions in these nuclei.

\section*{Acknowledgements}

This work is supported in part by the NSFC (12022504, 11775273, U1932140,12121005),
by the CAS Pioneer Hundred Talents Program,
by the CAS Open Research Project of large research infrastructures,
by the CAS Maintenance and Reform of large Research infrastructures (DSS-WXGZ-2018-0002),
and by the Max-Plank-Society.
Y.A.L. acknowledges the European Research Council (ERC) under the European Union’s Horizon 2020 research and innovation programme (grant agreement No 682841 “ASTRUm”) and
P.S. acknowledges MCI/AEI/FEDER, UE (Spain) (grant PGC2018-093636-B-I00).

\section*{Appendix}
{\scriptsize
\tablecaption{Experimental and evaluated neutron skin thicknesses. $\sigma_{I}$, AA, GDR, SDR, and PDR denote the interaction cross section, antiprotonic atom, giant dipole resonance, spin dipole resonance, and pygmy dipole resonance, respectively.}
\tablefirsthead{\hline\hline
\multicolumn{1}{c}{} &
\multicolumn{1}{c}{Experimental }&
\multicolumn{1}{c}{Method}&
\multicolumn{1}{c}{Evaluated}&
\multicolumn{1}{c}{Difference} \\
\multicolumn{1}{c}{} &
\multicolumn{1}{c}{${\Delta r}_{np}^{exp}$ (fm)}&
\multicolumn{1}{c}{}&
\multicolumn{1}{c}{${\Delta r}_{np}^{eva}$ (fm)}&
\multicolumn{1}{c}{$\frac{\Delta r_{np}^{eva}-\Delta r_{np}^{exp}}{error}$} \\
\hline }
\tablehead{\multicolumn{3}{c}{} \\
\hline\hline
\multicolumn{1}{c}{} &
\multicolumn{1}{c}{Experimental }&
\multicolumn{1}{c}{Method}&
\multicolumn{1}{c}{Evaluated}&
\multicolumn{1}{c}{Difference} \\
\multicolumn{1}{c}{} &
\multicolumn{1}{c}{${\Delta r}_{np}^{exp}$ (fm)}&
\multicolumn{1}{c}{}&
\multicolumn{1}{c}{${\Delta r}_{np}^{eva}$ (fm)}&
\multicolumn{1}{c}{$\frac{\Delta r_{np}^{eva}-\Delta r_{np}^{exp}}{error}$} \\
\hline}
\tabletail{%
\hline
\multicolumn{2}{c}{{}}\\}
\tablelasttail{%
\hline\hline
}
 \centering
\begin{xtabular}{ccccc}
 \label{tab1}

 $^{40}$Ca &-0.080(1000)~\cite{Trzcinska01,Jastrzebski04}&AA                       &0.007(13)   &0.1\\
                      &0.020(30)~\cite{Chaumeaux78}                         &(p,p)                     &                   &-0.4\\
                    &-0.070(120)~\cite{Brissaud72}                         &$(\alpha,\alpha)$ &                    &0.6\\
                    &-0.016(50)~\cite{Alkhazov77-1}                       &$(\alpha,\alpha)$ &                 &0.5 \\
                    &-0.009(140)~\cite{Papanicolas82}                    &$(\alpha,\alpha)$ &                 & 0.1\\
                    &-0.009(40)~\cite{Gils84}                                   &$(\alpha,\alpha)$ &                 &0.4\\
                   &0.010(140)~\cite{Lombardi72}                          &(p,p)                     &                 &0.0\\
                   &0.000(60)~\cite{Alkhazov75}                              &(p,p)                     &                 &0.1\\
                   &0.014(30)~\cite{Alkhazov76}                             &(p,p)                      &                 &-0.2\\
                   &-0.070(50)~\cite{Varma77}                                 &(p,p)                      &                  &1.5\\
                  &0.100(50)~\cite{Ray79}                                       &(p,p)                      &                 &-1.9\\
                  &0.010(80)~\cite{Igo79}                                        &(p,p)                     &                 &0.0\\
                  &-0.010(100)~\cite{Alkhazov82}                          &(p,p)                     &                 &0.2\\
                  &0.030(50)~\cite{McCamis86}                             &(p,p)                    &                   &-0.5\\
                  &-0.010(49)~\cite{Zenihiro18}                             &(p,p)                      &                  &0.3\\

\hline
 $^{42}$Ca &0.080(30)~\cite{Chaumeaux78}                      &(p,p)                       &0.055(16)  &-0.8 \\
                    &0.043(47)~\cite{Alkhazov77-1}                       &$(\alpha,\alpha)$ &                     &0.3\\
                    &-0.030(134)~\cite{Papanicolas82}                  &$(\alpha,\alpha)$  &                  &0.6\\
                    &0.027(38)~\cite{Gils84}                                   &$(\alpha,\alpha)$  &                  &0.7    \\
                    &0.055(30)~\cite{Alkhazov76}                          &(p,p)                       &                  &0.0  \\
                    &0.080(80)~\cite{Igo79}                                    &(p,p)                       &                  &-0.3  \\
                    &0.060(130)~\cite{McCamis86}                       &(p,p)                       &                  &0.0        \\
                    &0.049(60)~\cite{Tanaka20,Tagami20}             &$\sigma_{I}$        &                  &0.1\\
\hline
 $^{44}$Ca&0.130(30)~\cite{Chaumeaux78}                    &(p,p)                        & 0.091(15)  &-1.3 \\
                    &0.090(160)~\cite{Brissaud72}                        &$(\alpha,\alpha)$   &                    &0.0\\
                    &0.079(45)~\cite{Alkhazov77-1}                      &$(\alpha,\alpha)$   &                   &0.3 \\
                    &-0.011(129)~\cite{Papanicolas82}                 &$(\alpha,\alpha)$   &                    &0.8 \\
                    &0.044(36)~\cite{Gils84}                                 &$(\alpha,\alpha)$   &                    &1.3 \\
                    &-0.020(120)~\cite{Lombardi72}                    &(p,p)                       &                     &0.9 \\
                    &0.088(30)~\cite{Alkhazov76}                         &(p,p)                       &                     &0.1 \\
                    &0.100(80)~\cite{Igo79}                                  &(p,p)                         &                    &-0.1\\
                    &0.110(170)~\cite{McCamis86}                     &(p,p)                         &                    &-0.1\\
                    &0.125(50)~\cite{Tanaka20,Tagami20}          &$\sigma_{I}$          &                     &-0.7\\
\hline
 $^{46}$Ca&0.151(50)~\cite{Tanaka20,Tagami20}          &$\sigma_{I}$        &0.151(50)      &0.0\\
\hline
 $^{48}$Ca &0.090(50)~\cite{Trzcinska01,Jastrzebski04} &AA                       & 0.191(13)     &2.0   \\
                     &0.210(30)~\cite{Chaumeaux78}                  &(p,p)                          &                        &-0.6   \\
                    &0.330(120)~\cite{Brissaud72}                    &$(\alpha,\alpha)$     &                           &-1.2 \\
                    &0.196(42)~\cite{Alkhazov77-1}                  &$(\alpha,\alpha)$     &                        &-0.1 \\
                    &0.096(119)~\cite{Papanicolas82}               &$(\alpha,\alpha)$     &                         &0.8\\
                    &0.214(50)~\cite{Gils84}                               &$(\alpha,\alpha)$    &                         &-0.5 \\
                    &0.390(100)~\cite{Lombardi72}                  &(p,p)                           &                        &-2.0\\
                    &0.130(60)~\cite{Alkhazov75}                     &(p,p)                           &                        &1.0 \\
                    &0.190(30)~\cite{Alkhazov76}                      &(p,p)                          &                        &0.0 \\
                    &0.210(50)~\cite{Varma77}                           &(p,p)                          &                        &-0.4  \\
                    &0.230(50)~\cite{Ray79}                               &(p,p)                          &                       &-0.8 \\
                    &0.180(80)~\cite{Igo79}                                 &(p,p)                         &                       &0.1 \\
                    &0.160(100)~\cite{Alkhazov82}                    &(p,p)                          &                      &0.3 \\
                    &0.220(110)~\cite{McCamis86}                   &(p,p)                          &                       &-0.3 \\
                    &0.168(55)~\cite{Zenihiro18}                        &(p,p)                         &                        &0.4  \\
                    &0.146(60)~\cite{Tanaka20,Tagami20}          &$\sigma_{I}$          &                     &0.8   \\

\hline

 $^{58}$Ni  &-0.090(160)~\cite{Trzcinska01,Jastrzebski04}&AA                      &-0.008(17)                 &0.5\\
                       &-0.010(30)~\cite{Chaumeaux78}                   &(p,p)                           &                   &0.1\\
                    &0.010(100)~\cite{Brissaud72}                      &$(\alpha,\alpha)$      &                   &-0.2\\
                    &-0.097(137)~\cite{Papanicolas82}               &$(\alpha,\alpha)$       &                   &0.6 \\
                      &0.010(50)~\cite{Ray79}                                  &(p,p)                           &                  &-0.4  \\
                   &0.030(120)~\cite{Zamora17}                         &$(\alpha,\alpha)$      &                   &-0.3 \\
                   &0.180(200)~\cite{Greenlees70}                     &(p,p)                          &                   &-0.9 \\
                   &0.010(80)~\cite{Hoffmann78-1}                   &(p,p)                           &                   &-0.2 \\
                   &-0.011(30)~\cite{Blanpied77}                        &(p,p)                          &                    &0.1  \\
                   &-0.036(70)~\cite{Ray78}                                &(p,p)                           &                   &0.4 \\
                   &-0.010(100)~\cite{Lombard81}                      &(p,p)                           &                 &0.0  \\
                   &0.096(248)~\cite{Yue19}                                 &(p,p)                           &                &-0.4 \\

 \hline
 $^{60}$Ni&-0.010(150)~\cite{Trzcinska01,Jastrzebski04}&AA                          &-0.011(58)      &0.0\\
                    &0.080(100)~\cite{Brissaud72}                        &$(\alpha,\alpha)$      &                 &-0.9\\
                    &-0.051(132)~\cite{Papanicolas82}                 &$(\alpha,\alpha)$       &                 &0.3 \\
                    &-0.080(100)~\cite{Lombard81}                      &(p,p)                            &                 &0.7\\

\hline
 $^{62}$Ni&0.090(100)~\cite{Brissaud72}                         &$(\alpha,\alpha)$      &0.075(62) &-0.2\\
                   &0.044(127)~\cite{Papanicolas82}                      &$(\alpha,\alpha)$      &                  &0.2\\
                   &0.080(100)~\cite{Lombard81}                          &(p,p)                            &                 &-0.1 \\
\hline
 $^{64}$Ni &0.040(80)~\cite{Trzcinska01,Jastrzebski04} &AA                             &0.129(34)                  &1.1 \\
                    &0.100(123)~\cite{Papanicolas82}                    &$(\alpha,\alpha)$     &                       &0.2\\
                    &0.170(50)~\cite{Ray79}                                    &(p,p)                          &                  &-0.8  \\
                    &0.180(80)~\cite{Hoffmann78-1}                      &(p,p)                      &                  &-0.6   \\
                    &0.040(100)~\cite{Lombard81}                         &(p,p)                           &                  &0.9\\

\hline
$^{112}$Sn&0.070(20)~\cite{Trzcinska01,Jastrzebski04}  &AA             &0.070(20)  &0.0\\

\hline
 $^{114}$Sn&0.040(50)~\cite{Krasznahorkay99}                &SDR             &0.040(50)   &0.0\\
 \hline
 $^{116}$Sn  &0.110(18)~\cite{Terashima08}                      &(p,p)                           &0.111(13)   &0.1  \\
                      &0.100(30)~\cite{Trzcinska01,Jastrzebski04}  &AA              &                  &0.4 \\
                      &0.080(90)~\cite{Brissaud72}                         &$(\alpha,\alpha)$        &                    &0.3\\
                       &0.150(50)~\cite{Ray79}                                 &(p,p)                              &                  &-0.8  \\
                          &0.130(70)~\cite{Ray78}                                 &(p,p)                              &                  &-0.3  \\
                       &0.120(60)~\cite{Krasznahorkay99}              &SDR               &                  &-0.2 \\
                      &0.120(60)~\cite{Hoffmann78-2}                   &(p,p)                             &                   &-0.2 \\
                      &0.020(120)~\cite{Krasznahorkay94}            &GDR         &                   &0.8 \\

\hline
 $^{118}$Sn&0.145(16)~\cite{Terashima08}                     &(p,p)                                 &0.145(15)  &0.0 \\
                        &0.170(90)~\cite{Brissaud72}                       &$(\alpha,\alpha)$            &                &-0.3\\
                       &0.130(60)~\cite{Krasznahorkay99}             &SDR                   &                  &0.3 \\
\hline
 $^{120}$Sn   &0.147(33)~\cite{Terashima08}                   &(p,p)                                 &0.137(23)                    &-0.3  \\
                      &0.080(40)~\cite{Trzcinska01,Jastrzebski04} &AA               &                    &1.4  \\
                       &0.230(90)~\cite{Brissaud72}                      &$(\alpha,\alpha)$            &                &-1.0\\
                       &0.250(200)~\cite{Greenlees70}                &(p,p)                                 &                     &-0.6 \\
                       &0.180(60)~\cite{Krasznahorkay99}           &SDR                  &                    &-0.7 \\

\hline
 $^{122}$Sn&0.146(16)~\cite{Terashima08}                   &(p,p)                                 &0.151(15)                  &0.3\\
                       &0.220(70)~\cite{Krasznahorkay99}            &SDR                                 &                 &-1.0  \\
                      &0.200(90)~\cite{Mailandt73}                     &$(\alpha,\alpha)$             &       &-0.5\\
                       
\hline
 $^{124}$Sn &0.185(17)~\cite{Terashima08}                     &(p,p)                                &0.183(13)                  &-0.1 \\
                       &0.140(30)~\cite{Trzcinska01,Jastrzebski04}  &AA             &                  &1.4\\
                      &0.160(90)~\cite{Brissaud72}                      &$(\alpha,\alpha)$            &                  &0.3\\
                      &0.250(50)~\cite{Ray79}                               &(p,p)                                 &                 &-1.3\\
                     &0.220(70)~\cite{Ray78}                               &(p,p)                                 &                 &-0.5\\
                    &0.190(70)~\cite{Krasznahorkay99}              &SDR                &                 &-0.1 \\
                     &0.200(60)~\cite{Hoffmann78-2}                &(p,p)                                 &                 &-0.3 \\
                      &0.210(110)~\cite{Krasznahorkay94}           &GDR           &                 &-0.2 \\

\hline
$^{204}$Pb&0.220(90)~\cite{Gils76}                               &$(\alpha,\alpha)$          &0.191(49)   &-0.3\\
                     &0.178(59)~\cite{Zenihiro10}                         &(p,p)                              &                     &0.2\\
\hline
 $^{206}$Pb&0.190(90)~\cite{Gils76}                             &$(\alpha,\alpha)$           &0.182(34)    &-0.1\\
                       &0.180(64)~\cite{Zenihiro10}                       &(p,p)                               &                     &0.0         \\
                      &0.181(45)~\cite{Starodubsky94}                 &(p,p)                               &                     &0.0 \\
                     
\hline
 $^{208}$Pb&0.150(20)~\cite{Trzcinska01,Jastrzebski04}  &AA                            &0.167(11)             &0.9 \\
                       &0.250(90)~\cite{Brissaud72}                     &$(\alpha,\alpha)$             &                 &-0.9\\
                        &0.080(50)~\cite{Varma77}                          &(p,p)                                   &                   &1.7\\
                      &0.160(50)~\cite{Ray79}                               &(p,p)                                  &                 &0.1 \\
                      &0.060(100)~\cite{Alkhazov82}                    &(p,p)                                  &                &1.1 \\
                      &0.360(200)~\cite{Greenlees70}                 &(p,p)                                  &                  &-1.0\\
                       &0.180(70)~\cite{Ray78}                               &(p,p)                                  &                  &-0.2\\
                       &0.190(90)~\cite{Krasznahorkay94}             &GDR             &                &-0.3\\
                      &0.300(70)~\cite{Gils76}                             &$(\alpha,\alpha)$               &                 &-1.9 \\
                     &0.211(63)~\cite{Zenihiro10}                       &(p,p)                                  &                  &-0.7 \\
                      &0.197(42)~\cite{Starodubsky94}                 &(p,p)                                  &                &-0.7  \\
                      &0.260(130)~\cite{Bernstein72}                  &$(\alpha,\alpha)$             &                   &-0.7\\
                      &0.420(200)~\cite{Bernstein72}                  &$(\alpha,\alpha)$             &                    &-1.3 \\
                      &0.273(90)~\cite{Tatischeff72}                    &$(\alpha,\alpha)$             &                   &-1.2\\

                      &0.182(70)~\cite{Blanpied78}                      &(p,p)                                  &                   &-0.2 \\
                    
                      &0.140(40)~\cite{Hoffmann80}                     &(p,p)                                 &                  &0.7\\

                      &0.180(35)~\cite{Klimkiewicz07}                &PDR      &                 &-0.4 \\
                    
                      &0.120(70)~\cite{Csatlos03}                         &GDR             &               &0.7 \\
                      &0.160(45)~\cite{klos07}                               &AA                   &               &0.2 \\

                      &0.200(64)~\cite{Brown07}                           &AA                    &                 &-0.5\\

\hline
            $^{106}$Cd&0.100(140)~\cite{Trzcinska01,Jastrzebski04}    &AA&0.100(140)&0.0\\

\hline
 $^{110}$Cd&0.076(14)~\cite{Miller81}            &$(\alpha,\alpha)$ &0.076(14)&0.0\\
\hline
 $^{112}$Cd&0.074(14)~\cite{Miller81}            &$(\alpha,\alpha)$ &0.074(14)&0.0\\
\hline
 $^{114}$Cd&0.090(15)~\cite{Miller81}            &$(\alpha,\alpha)$ &0.090(15)&0.0\\
\hline
 $^{116}$Cd&0.150(40)~\cite{Trzcinska01,Jastrzebski04}  &AA&0.111(14)&-1.0         \\
                      &0.105(15)~\cite{Miller81}         &$(\alpha,\alpha)$ &&0.4\\

 \hline

\end{xtabular}

}

\end{document}